\documentclass[prl,aps,singlecolumn]{revtex4}

\usepackage{graphics}

\begin{document}

\title{Transverse coincidence-structures in spontaneous parametric down-conversion \\with orbital angular momentum: Theory}

\author{Geraldo A. Barbosa$^*$}
\affiliation{Northwestern University, Center for Photonic Communication and Computing, Department of Electrical Engineering
and Computer Science, 2145 N. Sheridan Road, Evanston, IL 60208-3118 / $^*${\small \em E-mail: g-barbosa@northwestern.edu}}

\date{10 April 2007}
\begin{abstract}

\newcommand{\be}{\begin{equation}}
\newcommand{\ee}{\end{equation}}
\newcommand{\bea}{\begin{eqnarray}}
\newcommand{\eea}{\end{eqnarray}}

Coincidence-structures in the transverse plane of Type-II spontaneous parametric down-conversion carrying orbital angular
momentum are obtained. Azimuthal symmetry breaking around the pump beam direction reveals itself on these quantum images.
Analytical expressions for the amplitude probability of the down conversion process are shown  including the nonlinear
polarizability components.
\end{abstract}

\maketitle

\section{Introduction}

Entanglement of photon states carrying orbital angular momentum (OAM)  is a new tool in quantum optics. These states have
been experimentally produced by Spontaneous Parametric Down-Conversion (SPDC) \cite{mair}. They were first predicted
\cite{hugo-barbosa} subjected to certain symmetry conditions that are still subject to some controversy. Experimentally,
these states can be created under the condition of a-posteriori or a-priori OAM imprinting by appropriated masks or filter
converters. For the a-posteriori imprinting, the pump beam does not need to carry OAM but an appropriate OAM mode converter
$l$ is inserted onto one of the down converted beams. The conjugate beam is expected to acquire an OAM of $-l$. For the
a-priori imprinting, the pump beam mode is set in a OAM state, say, $l$. However, as discussed in \cite{hugo-barbosa} the
initial OAM $l$ may or may not be transferred to  the SPDC. Discussions about if this transfer of OAM occurs or not and if
special conditions are needed to have OAM transfer has populated the recent literature sometimes with conflicting answers:
``yes'', ``no'', ``under certain conditions'' (See some references in \cite{sheng_quantph}). Sometimes, conclusions derived
from particular cases say, under conditions adequate for low wave vectors or paraxial cases (e.g., \cite{WallbornMonken}),
can be mistakenly taken as being general. Part of this non-uniform understanding on this subject derives from oversimplified
Hamiltonian or wave-states considered (simplified ``backbone'' models). The difficulties to derive general conclusions
taking into account realistic phase matching and light-matter coupling through the non-linear polarizability tensor
sometimes makes these time consuming tasks not appealing. Therefore, it is common that a single proportionality constant
replaces an involved angular dependence resulting in a non realistic description. Instead of being just technical details,
these more complex dependences may be crucial to a full understanding of these processes including entanglement between
conjugate photon pairs. Phase matching is also frequently assumed under simplified conditions not sufficient to treat in
detail pump modes with amplitudes more complex than a simple gaussian intensity profile.

Although not claiming a complete analysis of this problem, this paper does not take for granted many of these commonly
oversimplified assumptions. It tries to discuss the problem of transfer of OAM between the pump beam and the SPDC photons
with reasonable detail. Phase matching conditions dependent on $\chi^{(1)}$ are not oversimplified; also detailed is the
dependence of SPDC on $\chi^{(2)}$. This should provide the reader with enough material to help his own work or, at least,
to show fundamental elements necessary in this trade. It may also stimulate others to improve our understanding on this
area. Certainly, many future applications will demand a careful understanding the entanglement carried by these non-linear
processes beyond current treatments. Even some small immediate rewards can be obtained such as a method to obtain the
non-linear coefficients in $\chi^{(2)}$ by comparison of experiment and theory.

This paper was written trying to give a newcomer to this field a straightforward view of the involved elements starting from
the wave state amplitude describing SPDC. It describes all components involved in this amplitude and detail aspects of phase
matching and simplifications adopted including the description of phase matching in two steps (longitudinal and transverse
conditions). The non-linear tensor $\chi^{(2)}$ and the resulting non-linear polarizability of the medium are discussed,
including their transformations between crystal reference system and the laboratory reference system.
This way, the reader can dedicated himself to the main aspects presented instead of having to work out technical details.
Certainly, for some, most of the presented aspects may be trivial and can be skipped without problems. However, any
disagreement between views may result in an overall enlightenment for all.

\section{Wave state}

The Hamiltonian describing SPDC can be easily found in the literature (e.g., \cite{MandelWolf}). It describes free
propagating photons from a pump and from the down-conversion process that occurs due to the non-linear light-matter
interaction occurring in a ideally transparent medium (e.g, crystalline medium within the crystal band gap).
A pump photon will excite the non-linear medium through a very fast interaction with virtual electrons and decay {\em
either}  into a similar pump photon or into two conjugate photons--historically called signal and idler. These virtual
interactions as well as the propagation of the SPDC photons occur in the non-isotropic medium with specific symmetries
defined by the crystal class involved. Therefore, medium symmetries are built explicitly in  $\chi^{(2)}$ and implicitly in
$\chi^{(1)}$, defining the non-linear interaction and the light propagation in the medium. The Hamiltonian for these
processes usually neglect coupling to the lattice possibly intermediated by electrons. Although the fast interaction times
indicate that this coupling to the lattice should be negligible, this possibility should not be ruled out in general.

 At this point, one could remind the reader the uncertainty
relationship connecting angular momentum and phase uncertainties \cite{PeggPadgett}
 \begin{eqnarray}
 \label{Pegg-Padgett}
  \Delta L_z \Delta \phi \geq \frac{\hbar}{2}\left(1- 2 \pi P(\phi_0) \right)\:\:.
  \end{eqnarray}
   It states that a large
uncertainty in angle allows $L_z$ to have a small uncertainty. Consequently, in order to guarantee precision in $L_z$,
uncertainty in $\phi$ has to be maximum. Could the medium symmetries (built in $\chi^{(2)}$ and $\chi^{(1)}$) cause
restrictions on the photon interactions or propagation that could somewhat constrain the associated azimuthal angle $\phi$
and therefore do not allow its maximum uncertainty to be achieved? This problem has been discussed in Ref.
\cite{hugo-barbosa} but it will detailed here for clarity.

The interaction Hamiltonian $H_I$ gives the wave state  $\left| \psi \left( t\right) \right\rangle =
\exp\left[(-i/\hbar)\int_{t-\tau}^t H_I(\tau^{\prime})d\tau^{\prime} \right]| 0\rangle $, from which successful spontaneous
photon  conversion in first order is
\begin{eqnarray}
\label{state_vector_final_form} \left| \psi \left( t\right)
\right\rangle = \sum_{s ,s ^{\prime }} \int \!\! d^3k^{\prime
}\!\!\int \!\!d^3k  \:  F_{s ,s ^{\prime }}({\bf k},{\bf
k}^{\prime})\: \widehat{a}^{\dagger }\left( {\bf k},s \right)
\widehat{a}^{\dagger }\left( {\bf k}^{\prime },s ^{\prime }\right)
\left| 0\right\rangle\:\:.
\end{eqnarray}
\vskip-3mm $\!\!\!\!\!$The probability amplitude for signal and
idlers at $({\bf k},{\bf k}^{\prime})$ is $F_{s ,s ^{\prime }}({\bf
k},{\bf k}^{\prime})=A_{{\bf k},s;{\bf k}^{\prime},s^{\prime}}
\:l_E^{(*)}\!\left( \omega_{\bf{k}} \right) l_E^{(*)}\!\left(
{\omega_{\bf{k}}}^{\prime}\right)
 T(\Delta \omega) \:
\widetilde{\psi }_{lp}\left( \Delta {\bf k}\right)$.  $ A_{{\bf
k},s;{\bf k}^{\prime},s^{\prime}}=\: \chi_{1jk}^{(2)}\left[ \left(
{\bf e}_{{\bf k},s }\right) _j^{*} \left( {\bf e}_{{\bf k}^{\prime
},s ^{\prime }}\right)_k^{*}+ \left( {\bf e}_{{\bf k}^{\prime },s
^{\prime }}\right)_j^{*} \left( {\bf e}_{{\bf k},s
}\right)_k^{*}\right]$, $({\bf e}_{{\bf k}},{\bf e}_{{\bf
k}^{\prime}})$ are unitary polarization vectors for signal and idler
photons, $l_E^{(*)}\!\left( \omega \right)=-i\sqrt{\hbar \omega/2
\epsilon({\bf k},s)}$, $\widetilde{\psi }_{lp}\left( \Delta {\bf
k}\right) =\int_{V_I} d^3r\:\psi _{lp}\left( {\bf r}\right) \exp
\left( -i{\Delta }{\bf k}\cdot {\bf r} \right) $ and $\psi _{lp}$ is
the Laguerre-Gaussian pump field amplitude in ${\bf E}\left( r ,\phi
,z;t\right) = \psi _{lp}\left( {\bf r}\right) e^{i\left( k_Pz-\omega
_P t\right) } \hat{{\bf e}}_1$. $T(\Delta
\omega)$=$\exp\left[i\Delta \omega \:(t-\tau/2)\right]\:
\sin\left(\Delta \omega\: \tau/2\right)/\left(\Delta \omega
/2\right) $ is the time window function defining the $\Delta \omega$
range given the interaction time $\tau$, $\Delta \omega
=\omega_{\bf{k}} +{\omega_{\bf{k}}}^{\prime}-\omega _P$, $\Delta
{\bf k=k+k} ^{\prime }-{\bf k}_P$.
Symmetries associated with the crystalline medium influence $A_{{\bf k},s;{\bf k}^{\prime},s^{\prime}}\times \widetilde{\psi
}_{lp}\left( \Delta {\bf k}\right)$  through $\chi^{(2)}$, $\chi^{(1)}$ and the polarization vectors \cite{hugo-barbosa}.

It was derived in Ref. \cite{hugo-barbosa} that in order for a wave state of a one-photon field $|\psi(t)\rangle_1$ to be an
eigenfunction of $J_z$ (or $L_z$ for processes where $S_z=0$, as will be assumed here) it has to obey  $J_z
|\psi(t)\rangle_1 = l \hbar|\psi(t)\rangle_1$. This directly gave
\begin{eqnarray}
\label{phasephi}
 | \psi(t)\rangle_1=\sum_s\int d^3k \:g\!\left(k_{\rho},k_z,s;t \right) e^{i l \phi} a^{\dagger}({\bf
k},s)|0\rangle\:\:.
\end{eqnarray}
In this condition one can see that the azimuthal phase should occur only as a phase term in the wave state and not in its
amplitude. This guarantees that the magnitude square of the integrand  presents complete rotational symmetry in $\phi$ or,
equivalently, presents a complete uncertainty in $\phi$, as required by Eq. (\ref{Pegg-Padgett}). This condition has to be
applied both to signal and idler photons. This implies that for perfect OAM transfer from pump photons to SPDC photons, the
probability amplitude $F_{s ,s ^{\prime }}({\bf k},{\bf k}^{\prime})$ in Eq. (\ref{state_vector_final_form}) should contain
the azimuthal angles for signal and idler only as phase terms $e^{i m \phi}$ and $e^{i n \phi^{\prime}}$. As a linear
superposition of wave functions is also a wave function, superpositions of expressions similar to Eq.
(\ref{state_vector_final_form}) containing similar phase terms can also represent a valid solution. However, the
superposition coefficients should not depend on the azimuthal angles.

Starting from the accepted standard wave state for SPDC, one may expand the most relevant terms to compare with symmetry
restrictions imposed by Eq. (\ref{phasephi}) and verify if a specific SPDC process may or may not transfer OAM to the down
converted photons or even if this transfer can be partial.  This answer should not be ambiguous and the result obtained
should be general enough to allow direct comparison with specific experiments.

\section{The wave state amplitude  $F({\bf k}_s,{\bf k}_i)$}

The wave state amplitude  $F({\bf k}_s,{\bf k}_i)$ (see Eq. \ref{state_vector_final_form}) is the term that needs to be
considered in detail. In principle, this amplitude contains a wealthy of information one could obtain from the wave state,
from efficiency considerations to phase matching conditions. Of course, one has to assume complementary information about
the crystalline medium including the light propagation conditions given by Fresnel equations. From the wave state amplitude
the required conditions for phase matching emerge naturally.

The term $A_{{\bf k},s;{\bf k}^{\prime},s^{\prime}}$ will be considered in a separate section and the time window term $
 T(\Delta \omega)$ admits easy interpretations. For example, for a CW laser where the interaction time $\tau$ can be made large    $T
(\Delta \omega)\rightarrow \pi \delta(\Delta \omega)$. The most involved term is the Fourier transform ${\widetilde
\psi}_{lp}(\Delta {\bf k})$. A careful consideration of this term allows one to derive main conclusions about the SPDC
process in general.

\subsection{The Fourier transform ${\widetilde \psi}_{lp}(\Delta {\bf k})$}

The Fourier transform  ${\widetilde \psi}_{lp}(\Delta {\bf k})$ of $\psi_{lp}(\Delta {\bf k})$ can be written
\begin{eqnarray}
\label{int}{\widetilde \psi}_{lp}(\Delta {\bf k})=\int_{z_0-l_c/2}^{z_0+l_c/2} dz \int_{0}^{2 \pi}d\phi \int_{0}^{\infty}dr
\: \psi_{lp}(r,\phi,z)\:e^{-i  \left( \Delta k_x r\cos \phi +\Delta k_y r \sin \phi +\Delta k_z z\right)}\:\:,\end{eqnarray}
where
\begin{eqnarray} \label{LG2}
\psi_{lp}(r,\phi,z)=\frac{A_{lp}}{\sqrt{1+(z/z_R)^2}} \left[\frac{r\sqrt{2}}{w(z)} \right]^l \!\!L_p^l\left[\frac{2
r^2}{w(z)^2} \right] \! \exp\left[-i\left(\frac{k_P\:  r^2 z}{2q(z)}+l\tan^{-1}\frac{y}{x}
\right)\right]\!\exp\left[i(2p+l+1)\:\tan^{-1}\frac{z}{z_R} \right],
\end{eqnarray}
and $w^2=w_0^2 \left[1 +\left(z^2/z_R^2\right) \right],\:q(z)=z+i z_R,\:w_0^2=\left(2 z_R/k_P\right),$ $z_0$ gives the
crystal center, $r^2=x^2+y^2$, and $l_c$ is the crystal length along the propagation direction. Development of integral
(\ref{int}) in $r$ and $\phi$ is straightforward although cumbersome:
\begin{eqnarray}
&&{\widetilde \psi}_{lp}(\Delta {\bf k})= \pi A_{lp} \left(
\frac{i}{2} \right)^l  \left( \frac{z_R}{k_P}
\right)^{1+\frac{l}{2}} e^{-i l \pi/2} \rho_k^l \:L_p^l\left(
\frac{z_R}{k_P} \rho_k^2\right) e^{- z_R \rho_k^2/(2 k_P)} e^{il
\tan^{-1}(\Delta k_y/\Delta k_x)}\nonumber\\&& \times
\int_{z_0-l_c/2}^{z_0+l_c/2}
 e^{i \left( -1+l+2 p \right)\tan^{-1}(z_R/z)} e^{i(1+l+2 p)\tan^{-1}( z /(2
z_R))}  \:e^{-i \Delta k_z}dz\:\:,
\end{eqnarray}
where $\rho_k^2=\Delta k_x^2+\Delta k_y^2=\rho^2+\rho^{\prime 2}+2\rho \rho^{\prime} \cos(\phi-\phi^{\prime})$, $\rho=k \sin
\theta$ and $\rho^{\prime}=k^{\prime} \sin \theta^{\prime}$.

The remaining $z$-integral can be solved under conditions favorable to experiments. Conditions $z_0=0$ and $l_c \ll z_R$ are
most usual. This gives $\tan^{-1}(z/(2z_R))\rightarrow 0$ and $\tan^{-1}(z_R/z)\rightarrow \pi/2$; therefore
\begin{eqnarray}
\label{ampl} {\widetilde \psi}_{lp}(\Delta {\bf k})\!=\!
 -\pi i^{l+1} 2^{-l} \!A_{lp} \!
 \left(\! l_c \frac{z_R}{k_P}\!\right) \!e^{i \left[ \pi p+l \tan^{-1}(\Delta k_y/\Delta k_x) \right]}
 e^{-\xi/2}\xi^{l/2} L_p^l\left( \xi \right)\!
 \frac{\sin \left[\frac{l_c}{4 z_R}\left(4-\!2 z_R \Delta k_z \!+\! \xi \right) \right]}{
 \left[ \frac{l_c}{4 z_R}\left(4-\!2 z_R \Delta k_z\!+\!\xi \right)\right]}
 ,\mbox{where}\:\:\xi\equiv \!
\frac{z_R}{k_P}\rho_k^2\:.
\end{eqnarray}
It is interesting to observe the presence of the phase $\tan^{-1}(\Delta k_y/\Delta k_x)$ in $ {\widetilde \psi}_{lp}(\Delta
{\bf k})$. It gives several possibilities for entanglement of signal and idler phases $\phi$ and $\phi^{\prime}$ and is a
signature of the complexity connecting signal and idler photons on the plane transverse to the propagation direction. The
probability for unconstrained signal and idler occurrences $|F_{s ,s ^{\prime }}({\bf k},{\bf k}^{\prime})|^2$ will be
proportional to
\begin{eqnarray}
\label{prob} |{\widetilde \psi}_{lp}(\Delta {\bf k})|^2\!=\!\pi^2
4^{-l}  |A_{lp}|^2  \left( l_c \frac{z_R}{k_P}\right)^2\!
 e^{- \xi} \:\xi^{l}\: L_p^l\left(\xi\right)^2 \!
\left(\frac{\sin \left[\frac{l_c}{4 z_R}\left(4-\!2 z_R \Delta k_z
\!+\! \xi \right) \right]}{\left[ \frac{l_c}{4 z_R}\left(4-\!2 z_R
\Delta k_z\!+\!\xi \right)\right]}\right)^2\:\:.
\end{eqnarray}
 Eqs. (\ref{ampl}) and (\ref{prob}) are fundamental for SPDC. Together with $A_{{\bf k},s;{\bf k}^{\prime},s^{\prime}}$
 they determine
probabilities of signal and idler occurrences and, by state superpositions with arbitrary phases, interferences. Phase
matching are determined by the loci of maxima of these same equations.

\section{Phase matching}

Phase matching conditions with variables  $\Delta k_z$ and $\xi$ can be obtained from $|{\widetilde \psi}_{lp}(\Delta {\bf
k})|^2$. In principle, polar and azimuthal angles that define maxima of $|{\widetilde \psi}_{lp}(\Delta {\bf k})|^2$ should
be determined simultaneously but simplified solutions are frequently used according to the level of detail one requires. For
example, since $\sin x/x$  in Eq.~(\ref{prob}) is weakly dependent on $\xi$, Eq. (\ref{prob}) can be treated as two
independent parts for phase matching considerations. One part, in form of $\sin x/x$ gives the $\Delta k_z $ range and the
second or remaining part of the equation sets the width associated with the variable $\xi$. Write $|{\widetilde
\psi}_{lp}(\Delta {\bf k})|^2=f_{\mbox{\tiny long}}\times f_{\mbox{\tiny transv}}$, where  $f_{\mbox{\tiny long}}=(\sin
x/x)^2$ and $f_{\mbox{\tiny transv}}=\pi^2 4^{-l} |A_{lp}|^2  \left( l_c z_R/k_P\right)^2\!
 e^{- \xi} \:\xi^{l}\: L_p^l\left(\xi\right)^2$.
 $f_{\mbox{\tiny long}}$ can be considered non-negligible from
$\Delta k_{z\mbox{\tiny min}}=(4+\xi)/2 z_R\simeq 2/z_R$  to $\Delta k_{z\mbox{\tiny max}}=(l_c (4+\xi)-4 \pi z_R)/2 l_c
z_R$ (for $x=\pi$). For $l_c\ll 1$, $\Delta k_z=\pm 2\pi/l_c$. This sets the polar angles $\theta_{pm}$ and
$\theta_{pm}^{\prime}$ for phase matching. Maxima for $f_{\mbox{\tiny transv}}$ will describe azimuthal angles, $\phi$ and
$\phi^{\prime}$, and $\theta_{pm}$ and $\theta_{pm}^{\prime}$, that are the polar angle values defined from $\Delta
k_{z\mbox{\tiny min}}$ to $\Delta k_{z\mbox{\tiny max}}$. These maxima are not hard to find for specific values of $l$ and
$p$.

Just to exemplify a simplified use of Eq. (\ref{prob}), one can calculate the probability for signal and idler occurrences
in a specific Type II case of a uniaxial crystal where the pump beam carries a $l=4$ OAM. The calculations are done for a
BBO crystal with the crystal axis inclined by $\theta_c$ with respect to the pump beam propagation direction. Two natural
coordinate systems are involved, the crystal and the laboratory axes. While description of the linear and non-linear
susceptibilities are usually given in the crystal reference system, the pump beam and the characteristic SPDC pattern
desired define the laboratory system to be used. Appropriate coordinate transformations have to applied to provide correct
answers.
 Starting with the longitudinal equation for phase matching, some additional information needed for its solution will be
presented. These additional information will also be used for the transverse phase matching conditions.

\subsection{Longitudinal condition}

Expanding $\Delta k_z$ in the longitudinal condition gives the interval where an appreciable contribution is found: $\Delta
k_z=\pm 2\pi/l_c$. That is to say, for angles bounded by this condition, there is a good probability to find signal and
idler photons. Let us re-write this condition in a more tight bound and expand $\Delta k_z$:
\begin{eqnarray}
\label{longitudinal} -k_P+\left(   k \cos\theta+k^{\prime} \cos\theta^{\prime}\right) \simeq \frac{2 \pi}{l_c}\:\:.
\end{eqnarray}
The wave vectors in the medium are $k=(2\pi/\lambda)n=(\omega/c) n$ and
$k^{\prime}=(2\pi/\lambda^{\prime})n^{\prime}=(\omega^{\prime}/c) n^{\prime}$ where $\lambda$ and $\lambda^{\prime}$ are
vacuum wavelengths and the refractive indexes $n$ and $n^{\prime}$ have to be found using Fresnel's equations for specific
propagation directions and Sellmeier's equations for the principal refractive indexes.

\subsubsection{Fresnel's equations}

Fresnel's equations can provide the refractive index $n$ of a uniaxial crystal along an arbitrary propagation direction
specified by the wave vector ${\bf k}=(\omega n/c )\:\hat{{\bf s}}$
where the unit vector is $\hat{{\bf s}}=s_x \hat{{\bf x}}+s_y \hat{{\bf y}}+s_z \hat{{\bf z}}$.  They are derived \cite{BW}
starting from Maxwell's Eqs. written in local variables:
$  \hat{{\bf s}}\cdot{\bf H}=0\:\:,\:\:{\bf D}\cdot \hat{{\bf s}}=0\:\:,\:\: {\bf H}\times\hat{{\bf s}}={\bf D}/n\:\:,\:\:
{\bf E}\times\hat{{\bf s}}=-{\bf H}/n$,
where
${\bf \nabla}$ was written as ${\bf \nabla}\rightarrow i
 (\omega n/c)\hat{{\bf s}}$.
Eliminating ${\bf H}$ results $ {\bf D}_j= n^2 \left[ {\bf E}_j -\hat{{\bf s}}_j\left( \hat{{\bf s}}\cdot {\bf E} \right)
\right]$. Considering an uniaxial crystal ($\epsilon_{ij}=\epsilon_{ii}\delta_{ij}$) and multiplying both members by $s_j$
one obtains $ \epsilon_{jj}E_j- n^2 E_j= -n^2\left[ s_j\left( \hat{{\bf s}}\cdot {\bf E} \right) \right]$ that gives $1=-n^2
s_j^2/(\epsilon_{jj}-n^2)$. Subtracting 1 ($=s_j^2$) from both sides and writing $n^2=\epsilon$, Fresnels's Eqs. are
obtained:
\begin{eqnarray} \label{fresnel} \frac{s_x^2}{\frac{1}{\epsilon}-\frac{1}{\epsilon_x}}+
\frac{s_y^2}{\frac{1}{\epsilon}-\frac{1}{\epsilon_y}}+ \frac{s_z^2}{\frac{1}{\epsilon}-\frac{1}{\epsilon_z}}=0\:\:,
\end{eqnarray} where $s_x=\sin\theta \cos\phi$, $s_y=\sin\theta \sin\phi$, $s_z=\cos\theta$ and
$\epsilon_x=\epsilon_y=\epsilon_o$ are the ordinary dielectric constants along the principal axis and
$\epsilon_z=\epsilon_e$ is the extraordinary dielectric constant.
(\ref{fresnel}).
Using the notation $r_x=\epsilon/\epsilon_o=r_y$ and $r_z=\epsilon/\epsilon_e$ and multiplying both members
of  Eq. (\ref{fresnel}) by $(1-r_x)^2(1-r_z)^2$ one obtains \begin{eqnarray} (1-r_x)^2(1-r_z)^2 \left[ \frac{  \left( 1-r_z
\right) \left(s_x^2+ s_y^2 \right) +\left( 1-r_x \right) s_z^2      }{(1-r_x)(1-r_z)} \right]=0 \:\:. \end{eqnarray} This
equation defines that {\em either}
$(1-r_x)(1-r_z)=0$,
that gives either $\epsilon=\epsilon_o$ or $\epsilon=\epsilon_e$ (pure ordinary and extraordinary cases), {\em or}
$\left[ \left( 1-r_z \right)  \left(s_x^2+ s_y^2  \right) +\left( 1-r_x \right) s_z^2 \right]=0$.
This last equation can be written as
\begin{eqnarray} \label{mixed}
\frac{1}{\epsilon}=\frac{s_x^2+s_y^2}{\epsilon_e}+\frac{s_z^2}{\epsilon_o}\:\:. \end{eqnarray} These are the possibilities
for light propagation in  uniaxial crystals. Either pure ordinary or extraordinary propagation or a mixture of ordinary and
extraordinary light propagation. However, these equations define the refraction indexes in the crystal coordinate system.
Obtaining the refractive indexes on the laboratory coordinate system demands that rotations are introduced corresponding to
the geometrical situation adopted.

\subsubsection{Rotation matrices}

\begin{figure}
\centerline{\scalebox{0.35}{\includegraphics{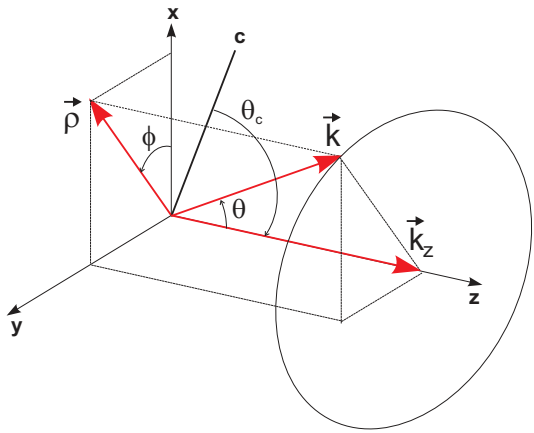}}} \caption{Decomposition of a wave vector ${\bf k}$ into
transversal $({\bf \rho})$ and longitudinal $({\bf k}_z)$ components in the laboratory axis $(x,y,z)$. A rotation around the
$y$-axis by $\theta_c$ separates the crystal principal axis from the pump beam propagation direction $z$.}
\label{LabCoordinates}
\end{figure}

One should be aware that rotation angles do not obey a universal notation and sometimes references for crystal rotation
angles may vary even from crystal to crystal in the literature. In particular, complementary angles may be referred to in a
similar way. Fig. \ref{LabCoordinates} shows a convenient laboratory coordinate system ($x,y,z$) obtained from the crystal
axis by rotation of angle $\theta_c$ around the $y$-axis.
Usual crystal rotations to utilize better geometries to increase the efficiency of the down-conversion process consist of
rotations about, say,  one of the crystal secondary axis followed by a rotation on its principal axis. These matrix
rotations can be written
\begin{eqnarray}
\label{Rzy} {\small R_{zy}=R_{z}(\phi_c)\!\cdot\!\! R_{y}(\theta_c)= \!\!\left(
\begin{array}{c c c}
\cos \phi_c &\sin \phi_c&0\\
-\sin \phi_c & \cos \phi_c &0\\
0&0&1
\end{array}  \right)\!\!\cdot\!\!
\left(
\begin{array}{c c c}
\cos \theta_c &0&-\sin \theta_c\\
0&1&0\\
\sin \theta_c& 0&\cos \theta_c
\end{array}  \right)
}\:\:.
\end{eqnarray}

The rotation matrix connecting crystal and laboratory coordinate systems adopted in this work is defined by Eq.~(\ref{Rzy}).
Of course, other commonly used rotation procedures exist. Rotation by two Euler angles, for example, are written not as
$R_{zy}$ but as $R_{yz}$, where the azimuthal rotation is applied first. These resulting rotations are not equal and they do
not have a unique correspondence using just two Euler angles. Although what description to adopt is a matter of taste, it
should be clearly stated to avoid misinterpretations about any result obtained.

\subsubsection{Refractive indexes} A wave-vector unitary propagation vector $\widehat{\bf k}$ written in the laboratory coordinates
$(\theta,\phi)$, $\widehat{\bf k}=(\sin \theta\cos \phi,\sin \theta \sin \phi,\cos \theta)$ is transformed to a vector
$\widehat{\bf s}$ in the crystal medium by $\widehat{\bf s}=R_{zy}\cdot \widehat{\bf k}$. The resulting pump, signal and
idler unitary wave vectors could then be plugged into Eq. (\ref{mixed}) to provide the corresponding refraction indexes in a
uniaxial crystal such as the BBO. One obtains
\begin{eqnarray}
n&=&n_{o}\\
 n^{\prime}&=&\frac{n^{\prime}_e n^{\prime}_o}{\sqrt{n^{\prime 2}_i \left(\cos \theta_c \cos \theta^{\prime}+\cos \phi^{\prime} \sin \theta_c \sin
\theta^{\prime}\right)^2+n^{\prime 2}_o \left[ \left(\cos\theta^{\prime}\sin \theta_c-\cos \theta_c \cos \phi^{\prime} \sin
\theta^{\prime}\right)^2 + \sin^2 \theta^{\prime} \sin^2 \phi^{\prime}\right]}}\\
n_P&=&\frac{n_{P,e}\:n_{P,o}}{\sqrt{n_{P,e}^2 \cos^2 \theta_c +n_{P,o}^2 \sin^2 \theta_c}}\:\:.
\end{eqnarray}
It is easy to observe that propagation under $eo$ polarizations do not present azimuthal symmetry. However, the overall
symmetry for SPDC depend on other terms as well.

\subsubsection{Sellmeier equations}

The refractive indexes in the principal axes,  $n_{P,e},n_{P,o},n_o,n_e,n^{\prime}_o$,  and $n^{\prime}_e$ are usually
obtained experimentally and represented by parametric equations representative of the microscopic physics involved. These
equations, known as Sellmeier's equations, have a wide applicability due to its success to represent the refraction index of
low absorption crystals as a real function of wavelength. They are based on the form of an oscillator response to an applied
force or, as a microscopic theory for the response of bound electrons in a solid to an applied electric field (See
\cite{BW}, \&2.3)
${\bf r}=e {\bf E} /\left[m \left( \omega_0^2 - \omega^2 \right) \right]$
or
$P=N e r \sim \chi^{(1)}E \sim ((n^2 -1)/4 \pi)E$
and assumes the form $ n^2=a+ b/(\lambda^2-c) -  d \lambda^2$. The constants $a$  to $d$ have to be experimentally
determined for each crystal along ordinary and extraordinary propagation directions and at every desired temperature. An
experimental fit to this  phenomenological equation may give a very good numerical representation of the refractive indexes
in a quite broad frequency range. The $d$ term is a first corrective term of a possible series of even terms in $\lambda$.
The assumption of reality for the fields ${\bf D}$ and ${\bf E}$ and their causal connection imposes the {\em even}
dependence on $\lambda$ for a real dielectric constant \cite{KramersKronig}. This way, the first terms are
\begin{eqnarray} n_o(\lambda)^2=a-\frac{b}{\lambda^2 -c}-d
\lambda^2\:\:,\:\:\mbox{and}\:\:n_e(\lambda)^2=e-\frac{f}{\lambda^2 -g}-h \lambda^2\:\:.
\end{eqnarray}

Ref. \cite{Eimerl_etal}, for example,  present the experimental values
$a=2.7405,\:b=0.0184,\:c=0.0179,\:d=0.0155,\:e=2.3730,\:f=0.0128,\:g=0.0156,\:h=0.0044$ to represent BBO's  refractive
indexes along the crystal axes ($\lambda$ are given in $\mu$m in these equations)

\subsubsection{Longitudinal condition} As the refractive indexes are now defined and the wave vectors $k=(2\pi
/\lambda)n$ can be calculated for general angles with adequate Sellmeier parameters, Eq. (\ref{longitudinal})
\begin{eqnarray}
-k_P+\left(   k \cos\theta+k^{\prime} \cos\theta^{\prime}\right) \simeq \frac{2 \pi}{l_c}\nonumber
\end{eqnarray}
can be solved. Due to the complexity dependence of the refractive indexes in the angles, this equation may not be solved
analytically in a general case. Numerical solutions can be found and will determine the angles where the SPDC process is
more efficient.
\begin{figure} \centerline{\scalebox{0.4}{\includegraphics{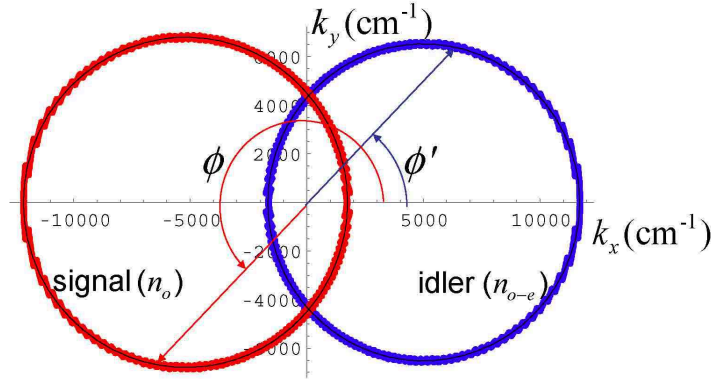}}} \caption{Signal and idler rings
obtained from numerical solutions and thin solid lines obtained from fitting the numerical equations with
Eqs.~(\ref{polarangles1}) and (\ref{polarangles2}).} \label{rings}
\end{figure}For a Type II process in BBO, for example, Fig.~\ref{rings} shows a plot obtained from the numerical solutions obtained. The obtained numerical solutions can be even
parameterized for simplicity. This way, polar angles giving SPDC rings are closely represented by
\begin{eqnarray}
\label{polarangles1} \theta_s = \!\arcsin\!\!\left[\zeta\left(\cos (\phi_{s}-\pi)+ \sqrt{ \cos^2 (\phi_{s}-\pi)+\eta
}\:\right)+\nu \exp\left( -\nu \sin^2\left[ (\phi_{s}-\pi)/2\right] \right)\cos\left[ 2(\phi_{s}-\pi) \right]
\right]\:,\end{eqnarray} and
\begin{eqnarray}\label{polarangles2} \theta_i=\!\arcsin\!\!\left[\zeta \! \left(\cos \phi_{i}+
\! \sqrt{ \cos^2 \phi_{i}+\eta }\,\right)\!+\!\nu \exp\left( -\nu \sin^2\left( \phi_{i}/2\right) \right)\cos 2\phi_{i}
\right]\end{eqnarray} where ${\small \zeta \simeq 0.034,\eta\simeq0.797,\nu\simeq0.0016,\mu\simeq 3.45}$. These equations
give the thin black lines in Fig.~\ref{rings}.

It has to be emphasized that these angles are described within the medium. Straightforward application of Snell's law gives
the angles outside of the crystal.

\subsection{Transverse equations}

The equation $f_{\mbox{\tiny transv}}=\pi^2 4^{-l}  |A_{lp}|^2  \left( l_c z_R/k_P\right)^2\!
 e^{- \xi} \:\xi^{l}\: L_p^l\left(\xi\right)^2$ together with the obtained polar angles given by Eqs. (\ref{polarangles1})
 and (\ref{polarangles2})
 define the loci of possibly entangled azimuthal angles that maximize $F_{s ,s ^{\prime }}({\bf
k},{\bf k}^{\prime})$. Finding the analytical maxima for $f_{\mbox{\tiny transv}}$ is not a trivial task and will not be
attempted here. However, for specific values of $p$ and $l$ this is usually a simple task.
\begin{figure}
\centerline{\scalebox{0.4}{\includegraphics{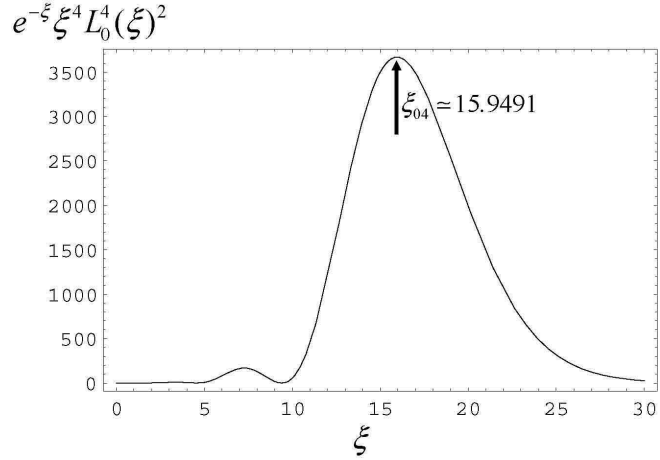}}} \caption{$e^{-\xi}\xi^4 L_0^4(\xi)^2$ as a function of $\xi$.}
\label{TransvMax}
\end{figure}
For example, Fig. \ref{TransvMax} shows a plot of $f_{\mbox{\tiny transv}}/(\pi^2 4^{-l}  |A_{lp}|^2  \left( l_c
z_R/k_P\right)^2)$ for $p=0$ and $l=4$. It is easy to find the maximum at $\xi_{04}\simeq15.9491$. This defines the value
$\rho_k^2(p=0,l=4)=(k_P/zR)\xi_{04}$ that maximizes the signal and idler emissions. From the transverse equation
\begin{eqnarray}
\label{transvcondition} \rho_k^2=\Delta k_x^2+\Delta k_y^2=\rho^2+\rho^{\prime 2}+2\rho \rho^{\prime}
\cos(\phi-\phi^{\prime})\simeq \frac{k_P}{z_R}\:\xi_{04}\:\:,
\end{eqnarray}
and using the polar angle dependence given by Eqs.~(\ref{polarangles1}) and (\ref{polarangles2}) this equation can be solved
for the azimuthal angles $\phi$ and $\phi^{\prime}$.
\begin{figure} \centerline{\scalebox{0.4}{\includegraphics{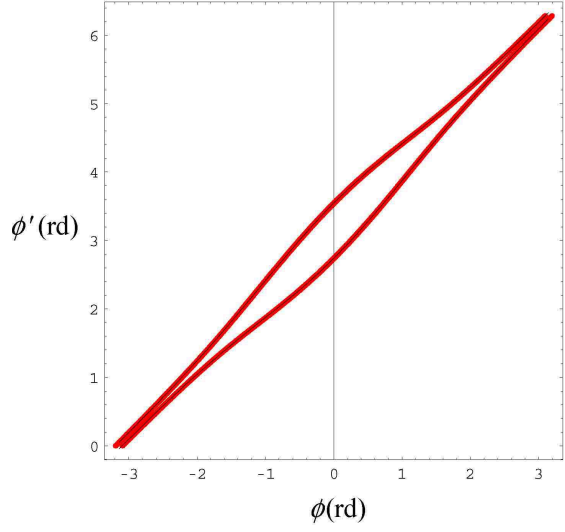}}}
\caption{Idler azimuthal angle versus signal azimuthal angle obtained numerically from Eq.~(\ref{transvcondition}). The thin
solid lines were obtained from Eqs.~\ref{transvfit}.} \label{phipVphi}
\end{figure}
Fig.~\ref{phipVphi} shows the numerical dependence found between these angles. While the condition $\phi^{\prime}=\phi+\pi$
is an approximated one, some deviations exist that may be meaningful in some applications.
These deviations are more apparent near azimuthal angles close to zero. The thin solid lines in Fig. \ref{phipVphi} were
obtained by fit, giving Eqs.~(\ref{transvfit}).
\begin{eqnarray}
\label{transvfit}
\phi_+^{\prime}&=&3.1836+ 0.3631 e^{-0.4653 \phi^2}+0.9999 \phi \nonumber \\
\phi_-^{\prime}&=&3.0996- 0.3631 e^{-0.4653 \phi^2}+0.9999 \phi \:\:.
\end{eqnarray}
The average azimuthal angle from these two equations is
\begin{eqnarray}
\phi^{\prime}=\frac{\phi_+^{\prime}+\phi_-^{\prime}}{2}=\phi+\pi\:\:.
\end{eqnarray}

\begin{figure}
\centerline{\scalebox{0.6}{\includegraphics{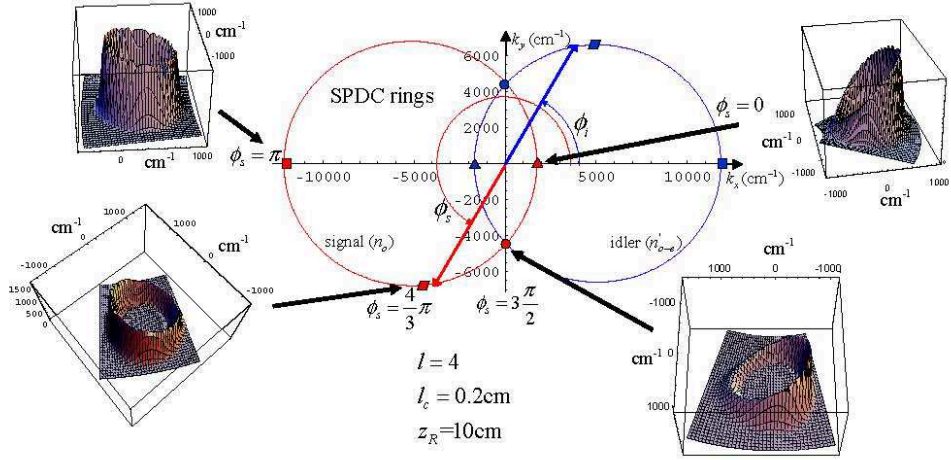}}} \caption{Calculated transverse coincidence-count structures
on s-ring with point detector on i-ring. All angles are lab angles but inside the crystal. The crystal is tilted with the
laser beam at $\theta_c=49.7^0$ ($\phi_c=0$) from the crystal c-axis. The laser wavelength is $\lambda_P=3511\AA(\hat{{\bf
e}}_1= \hat{{\bf x}}))$ and the principal refractive indexes are
$n_{P,o}\!=\!1.707,n_{P,e}\!=\!1.578,n_{o}\!=\!n^{\prime}_{o}\!=\!1.665,n^{\prime}_{e}\!=\!1.548$.} \label{TheoryTypeII}
\end{figure}

Ref.~ \cite{hugo-barbosa} pointed out that perfect transfer of OAM from the pump to the SPDC photons demands azimuthal
symmetry for $|F_{s ,s ^{\prime }}({\bf k},{\bf k}^{\prime})|^2$ around the pump propagation direction (quantization axis).
Lack of azimuthal symmetry causes partial transfer of OAM. To illustrate this partial transfer, Fig. \ref{TheoryTypeII}
shows transverse coincidence structures expected for degenerate non-collinear Type-II SPDC in a BBO crystal (See
\cite{ShengCharlie} for a Type-II experiment). Calculated structures represent the detection probability for signal and
idler photons within small $\Delta \omega \Delta\omega^{\prime} \Delta\theta \Delta\theta^{\prime}\Delta \phi \Delta
\phi^{\prime}$ around phase matching conditions (Here $\sin x/x \rightarrow 1$ and $dk \simeq (n/c)d\omega$)):
\begin{eqnarray}
\label{P} P_{\mbox{\tiny scatt}} \simeq \frac{1}{  | A_{{\bf
k},s;{\bf k}^{\prime},s^{\prime}}|^2}\times \left|\frac{d^3k
d^3k^{\prime }}{ d \omega d \omega^{\prime} d\theta
d\theta^{\prime}d \phi d \phi^{\prime}} \: F_{s ,s ^{\prime }}({\bf
k},{\bf k}^{\prime}) \right|^2 \:\:\:\left(\mbox{\small excluding
existing geometric effects given by}\:\:A_{{\bf k},s;{\bf
k}^{\prime},s^{\prime}}\right)\:.
\end{eqnarray}
The lack of azimuthal symmetry in Type-II SPDC is reflected on the coincidence structures (even with the neglect of
$|A_{{\bf k},s;{\bf k}^{\prime},s^{\prime}}|^2$) that show highly asymmetric coincidence donut like pattern.  Asymmetric
structures should then be expected in Type-II with OAM.  The $A_{{\bf k},s;{\bf k}^{\prime},s^{\prime}}$ contribution (to be
shown ahead) is non-negligible but presents no sharp variations for the structures.

\section{Non-linear polarizability}

The SPDC efficiency is directly proportional to the non-linear polarizability vector ${ \bf P}=\sum_i A_{i,({\bf k},s;{\bf
k}^{\prime},s^{\prime})} \widehat{x}_{i,cr}$ with components $ A_{i,({\bf k},s;{\bf k}^{\prime},s^{\prime})}$ given by the
product of the tensor $\chi_{1jk}^{(2)}$ and components of the unitary polarization vectors
\begin{eqnarray}
\label{A} A_{i,({\bf k},s;{\bf k}^{\prime},s^{\prime})}=\: \chi_{ijk}^{(2)}\left[ \left( {\bf e}_{{\bf k},s }\right) _j^{*}
\left( {\bf e}_{{\bf k}^{\prime },s ^{\prime }}\right)_k^{*}+ \left( {\bf e}_{{\bf k}^{\prime },s ^{\prime }}\right)_j^{*}
\left( {\bf e}_{{\bf k},s }\right)_k^{*}\right]\:\:.
\end{eqnarray}
The interaction energy $V_I$ is given by the product of the laser field polarization and the non-linear polarizability
vector, $V_I={\bf E}_{P}.{\bf  P}$. Up to know, all equations have been developed using laboratory coordinates. In order to
keep the same reference system, all quantities in $A_{{\bf k},s;{\bf k}^{\prime},s^{\prime}}$ have to be referred to the
laboratory coordinate system. Usually, the non-linear dielectric tensor is given in the crystal axes and a coordinate
transformation to the laboratory axes is necessary. $A_{{\bf k},s;{\bf k}^{\prime},s^{\prime}}$ involves the non-linear
tensor $\chi^{(2)}$ and the unitary polarization vectors ${\bf e}_{{\bf k},s }$. These constitutive elements will be
considered in the next sections. The non-linear tensor can be written in the laboratory coordinate system and multiplied by
the components of the unitary polarization vector in the same reference system or, alternately, $\chi_{ijk}^{(2)}$ and the
unitary polarization vectors can be written in the crystal reference system and the resulting vector component rotated to
the laboratory coordinate system. This last method will be followed.


The susceptibility tensor ${\chi}_{qmn}^{(2)}$ for the class of uniaxial crystals can be written in a contracted form as
${\chi}_{qmn}^{(2)}\rightarrow{\chi}_{ql}^{(2)}\equiv 2 d_{ql}$ as  indicated in Table~ \ref{contractions}.
\begin{table}
  \centering
  \begin{tabular}
{|c|cccccc|}\hline  {\bf indexes} & &&&{\bf contractions} && \\ \hline\hline
 {\bf \em  mn} & 11 & 22 & 33 & 23,32 & 31,13 & 12,21 \\\hline
 {\bf \em  l }& 1 & 2 &3 & 4 & 5 & 6 \\ \hline
\end{tabular}
  \caption{Contractions for tensor indexes.}\label{contractions}
\end{table}
This way,
\begin{eqnarray}
 \chi^{(2)}=
           2      \left(\begin{array}{cccccc}
                        d_{11} & d_{12}& d_{13}&d_{14}&d_{15}&d_{16}\\
                        d_{21} & d_{22}& d_{23}&d_{24}&d_{25}&d_{26}\\
                        d_{31} & d_{32}& d_{33}&d_{34}&d_{35}&d_{36}
\end{array} \right)
\end{eqnarray}


To calculate the unitary polarization vectors, one may write the electric displacement ${\bf D}$  in the crystal principal
axis, where $\epsilon_{ij}=\epsilon_{ii} \delta_{ij}$ results $D_i=\epsilon_{ii}\delta_{ii} E_j=n^2 \left[ E_i-s_i s_k E_k
\right]$ or
\begin{eqnarray}\label{fields}
\left[ n_x^2 - n^2 \left( 1-s_x^2 \right)\right]E_x+n^2 s_x s_y E_y + n^2 s_x s_z E_z&=&0 \nonumber \\
n^2 s_x s_y E_x + \left[ n_y^2-n^2 \left( 1-s_y^2 \right) \right]E_y+n^2 s_y s_ z E_z&=&0 \nonumber \\
n^2 s_x s_z E_x + n^2 s_y s_z E_y + \left[ n_z^2 - n^2 \left( 1-s_z^2 \right) \right]E_z&=&0\:\:. \end{eqnarray}
 This set of
Eqs. give the possible electric field amplitudes for specific propagation directions  and refractive indexes $n$. Without
developing general solutions, one may look for a solution for the signal photons propagating with {\em ordinary} refraction
index $n_o$. Writing $n_x=n_y=n_o$ and $n_z=n_e$ for the idler's refractive indexes,  the electric field components in the
medium are obtained and normalized resulting in
\begin{eqnarray}
\label{ecr} \widehat {\bf e}_{o,cr}=\left( -\sin \theta_{cr},\cos \theta_{cr},0  \right)\:\:.
\end{eqnarray}

To calculate the  unitary polarization vector for {\em extraordinary propagation}, one may observe that in the crystal
medium the Poynting vector form a set of orthogonal axes with the electric field and the magnetic field. At the same time,
the electric displacement may not be along with the electric field but is normal to the propagation vector in the medium.
Eq.~(\ref{fields}) can be particularized for this case as well. However, for simplicity, one may look for field solutions
that give a unitary polarization vector for the extraordinary propagation $\widehat {\bf e}_{cr}^{\prime}$ orthogonal both
to $\widehat {\bf e}_{cr}$ and to its unitary propagation vector $\hat{{\bf s}}^{\prime}=s_x^{\prime} \hat{{\bf
x}}+s_y^{\prime} \hat{{\bf y}}+s_z^{\prime} \hat{{\bf z}}$. This gives
\begin{eqnarray}
\label{epcr} \widehat {\bf e}_{e,cr}^{\prime}=\left(  -\cos \theta_{cr}^{\prime} \cos \phi_{cr}^{\prime}, -\cos
\theta_{cr}^{\prime} \sin \phi_{cr}^{\prime},\sin \theta_{cr}^{\prime} \cos \phi_{cr}^{\prime}\right)\:\:.
\end{eqnarray}

\subsection{Non-linear polarizabilities in the crystal reference system}

Calculation of the  non-linear polarizability components $A_{i,({\bf k},s;{\bf k}^{\prime},s^{\prime})}$ demands
specification of $\left[ \left( {\bf e}_{{\bf k},s }\right) _j^{*} \left( {\bf e}_{{\bf k}^{\prime },s ^{\prime
}}\right)_k^{*}+ \left( {\bf e}_{{\bf k}^{\prime },s ^{\prime }}\right)_j^{*} \left( {\bf e}_{{\bf k},s
}\right)_k^{*}\right]$ for all SPDC possible cases. The question mark in Eq.~(\ref{pol}) represents the need for these
choices. The possible cases $(o,o)$, $(e,e)$ , $(o,e)$ and $(o,e)$ are detailed in the Appendix.
\begin{eqnarray} \label{pol} {\bf P}=\left(\begin{array}{c} A_{1,({\bf
k},s;{\bf k}^{\prime},s^{\prime})} \\ A_{2,({\bf k},s;{\bf k}^{\prime},s^{\prime})} \\ A_{3,({\bf k},s;{\bf
k}^{\prime},s^{\prime})}  \end{array}  \right)=
           4      \left(\begin{array}{cccccc}
                        d_{11} & d_{12}& d_{13}&d_{14}&d_{15}&d_{16}\\
                        d_{21} & d_{22}& d_{23}&d_{24}&d_{25}&d_{26}\\
                        d_{31} & d_{32}& d_{33}&d_{34}&d_{35}&d_{36}
\end{array} \right)
                   \left(  \begin{array}{c}
                         e_{1}e_{1}^{\prime} \\ e_{2} e_{2}^{\prime}\\e_{3}e_{3}^{\prime}\\
       e_{2}e_{3}^{\prime}+e_{3}e_{2}^{\prime}\\  e_{1}e_{3}^{\prime}+e_{3}e_{1}^{\prime}\\  e_{1}e_{2}^{\prime}+e_{2}e_{1}^{\prime}
\end{array}  \right)_{?}\:\:.
\end{eqnarray}

Using the appropriate polarization cases the signal and idler non-linear polarizability  ${\bf P}_{oo}$, ${\bf P}_{ee}$,
${\bf P}_{eo}$ and ${\bf P}_{oe}$ are calculated (see Appendix). These non-linear polarizabilities cover all possibilities
of light excitation in uniaxial crystals.

\subsection{Non-linear polarizabilities in the laboratory reference system}

For SPDC it is usual to have the crystal rotated by convenient angles referred to the laboratory reference system
$x,y,x,\theta,\phi$  instead of $x_{cr},y_{cr},x_{cr},\theta_{cr},\phi_{cr}$. A laser with amplitude $E_P$ polarized along
direction $\widehat{\bf e}$ will define the light matter interaction to be analyzed. In this work it is chosen to have the
non-linear polarizability ${\bf P}$ rotated to the laboratory reference system; this transforms ${\bf P}\rightarrow {\bf
P}_{lab}$.

Eqs.~(\ref{Poo},\ref{Pee},\ref{Peo},\ref{Poe}) can be written in the laboratory coordinate system under rotation given by
$R_{zy}^{-1}$; ${\bf P}_{\alpha \beta;lab}=R_{zy}^{-1}.{\bf P}_{\alpha\beta},\:\:(\alpha\beta=oo,ee,eo,oe)$. For an example,
one can choose a laser amplitude ${\bf E}_P = E_P \widehat{\bf x}$ and obtain for ${o,e}$ polarizability
\begin{eqnarray}
V_{I,oe}=-E_P \widehat{\bf x} \cdot \left( R_{zy}^{-1}.{\bf P}_{oe}\right)=
 \cos {\theta_c}\,\cos {\phi_c}\,
     \left( - {d_{14}}\,\cos \phi_{cr}\,\sin \theta_{cr}^{\prime}   +
       {d_{12}}\,\cos \theta_{cr}^{\prime}\,\cos \phi_{cr}\,\sin \phi_{cr}^{\prime} \hspace{53cm}  \right. \nonumber\\ \left. -
       {d_{11}}\,\cos \theta_{cr}^{\prime}\,\cos \phi_{cr}^{\prime}\,\sin \phi_{cr} +
       {d_{15}}\,\sin \theta_{cr}^{\prime}\,\sin \phi_{cr} +
       {d_{16}}\,\left( \cos \theta_{cr}^{\prime}\,\cos \phi_{cr}^{\prime}\,\cos \phi_{cr} -
          \cos \theta_{cr}^{\prime}\,\sin \phi_{cr}^{\prime}\,\sin \phi_{cr} \right)  \right) \hspace{50cm} \nonumber\\ -
    \cos {\theta_c}\,\sin {\phi_c}\,
     \left( - {d_{24}}\,\cos \phi_{cr}\,\sin \theta_{cr}^{\prime}   +
       {d_{22}}\,\cos \theta_{cr}^{\prime}\,\cos \phi_{cr}\,\sin \phi_{cr}^{\prime} -
       {d_{21}}\,\cos \theta_{cr}^{\prime}\,\cos \phi_{cr}^{\prime}\,\sin \phi_{cr} \hspace{52cm}  \right. \nonumber\\ \left.+
       {d_{25}}\,\sin \theta_{cr}^{\prime}\,\sin \phi_{cr} +
       {d_{26}}\,\left( \cos \theta_{cr}^{\prime}\,\cos \phi_{cr}^{\prime}\,\cos \phi_{cr} -
          \cos \theta_{cr}^{\prime}\,\sin \phi_{cr}^{\prime}\,\sin \phi_{cr} \right)  \right)  \hspace{54.5cm}   \nonumber\\   +
    \sin {\theta_c}\,\left( -\left( {d_{34}}\,\cos \phi_{cr}\,\sin \theta_{cr}^{\prime} \right)  +
       {d_{32}}\,\cos \theta_{cr}^{\prime}\,\cos \phi_{cr}\,\sin \phi_{cr}^{\prime} -
       {d_{31}}\,\cos \theta_{cr}^{\prime}\,\cos \phi_{cr}^{\prime}\,\sin \phi_{cr}        \hspace{52.5cm}  \right. \nonumber\\ \left.      +
       {d_{35}}\,\sin \theta_{cr}^{\prime}\,\sin \phi_{cr}                 +
       {d_{36}}\,\left( \cos \theta_{cr}^{\prime}\,\cos \phi_{cr}^{\prime}\,\cos \phi_{cr} -
          \cos \theta_{cr}^{\prime}\,\sin \phi_{cr}^{\prime}\,\sin \phi_{cr} \right)  \right)\hspace{54.5cm}
\end{eqnarray}
For a crystal where the dominant coefficients are $d_{11}$, $d_{22}$ and $d_{15}$ (e.g. BBO), $V_{I,oe}$ simplifies to
\begin{eqnarray}
\label{Voe} V_{I,oe}= -\left({d_{22}}\cos {\theta_c}\cos \theta_{cr}^{\prime}\cos {\phi_{cr}}\,
       \sin {\phi_c}\sin \phi_{cr}^{\prime} \right)  +
    \cos {\theta_c}\cos {\phi_c}
     \left( - {d_{11}}\cos \theta_{cr}^{\prime}\cos \phi_{cr}^{\prime}\sin {\phi_{cr}}   +
       {d_{15}}\sin \theta_{cr}^{\prime}\sin {\phi_{cr}} \right)\:.
\end{eqnarray}
 Replacing $\theta_{cr}^{\prime}\rightleftharpoons \theta_{cr}$
$\phi_{cr}^{\prime} \rightleftharpoons \phi_{cr}$ in Eq.~(\ref{Voe}) gives $V_{I,eo}$ for BBO.

\subsubsection{Crystal to laboratory reference system}
If one wishes to have $V_{I,oe}$ (or any other interaction energy term) written under laboratory angles $(\theta,\phi)$, the
connection between angles in these two systems have to be found. Given a unitary vector ${\bf v}=\left( \sin \theta
\cos\phi,\sin \theta \sin \phi ,\cos \theta \right)$ in the laboratory reference system, the rotation $R_{zy}$ brings it to
the crystal coordinate system (or the inverse rotation to move the reference system). Assuming that vector components in the
crystal reference system can be written as ${\bf v}={\bf v}_{cr}=\left( \sin \theta_{cr} \cos\phi_{cr},\sin \theta_{cr} \sin
\phi_{cr} ,\cos \theta_{cr} \right)$, a connection between the medium and the laboratory angles can be found. This way
\begin{eqnarray}
 \theta_{cr}&=&\arccos\left[\cos \theta_c\,\cos \theta +
   \cos \phi\,\sin \theta_c\,\sin \theta\right]\\
\phi_{cr}&=&\arccos\left[ \frac{{\sqrt{{n1} + {n2} + {n3}}}}
   {{\sqrt{\left( -1 + \cos \theta_c\,\cos \theta +
          \cos \phi\,\sin \theta_c\,\sin \theta \right) \,
        \left( 1 + \cos \theta_c\,\cos \theta +
          \cos \phi\,\sin \theta_c\,\sin \theta \right) }}}\right]\:,
\end{eqnarray}
where
\begin{eqnarray}
n1&=&-1 + {\cos \phi}^2\,{\sin \theta_c}^2\,{\sin \theta}^2 +
  {\cos \theta_c}^2\,\left( {\cos \theta}^2 +
     {\cos \phi}^2\,{\sin \theta}^2\,{\sin \phi_c}^2 \right)\\
n2&=&{\left( \cos \theta \,\sin \theta_c\,\sin \phi_c +
     \cos \phi_c\,\sin \theta \,\sin \phi  \right) }^2 \\
n3&=&2 \cos \theta_c\,\cos \phi_c\,\cos \phi\,\sin \theta\,
  \left( \cos \theta\,\cos \phi_c\,\sin \theta_c -
    \sin \theta\,\sin \phi_c\,\sin \phi \right)\:.
\end{eqnarray}

Replacing in $V_{I,oe}$  the appropriate angle given by the $\theta_{cr}$ or $\phi_{cr}$ above, $V_{I,oe}$  is determined in
terms of the laboratory angles. The procedure is straightforward but the resulting expression is quite long.

\subsection{A method for determination of non-linear coefficients}

Knowing the precise rotation angles between the crystal reference system, the laboratory system and phase matching angles,
it was shown that all elements to determine the signal and idler scattering probabilities (wave state amplitude,
Eq.~(\ref{ampl})) can be obtained in detail. From these information, transverse coincidence-structures can be calculated
(see \cite{ShengCharlie}) and by integration over the angle variables for signal (or idler) the angular dependence of the
idler (or signal) intensity can be obtained. Aside from obtaining a detailed picture of SPDC process, it should be pointed
out that use of these equations allows one to obtain directly through simple intensity measurements \cite{Otavia} the
non-linear coefficients $d_{ij}$. This can be done by fit of the existing experimental angle dependence to the theoretical
elements presented, including numerical integrations indicated. Although this analysis is not the object of this work, it is
interesting to observe that considering just an integration over the nonlinear polarizability, neglecting the contribution
from $ {\widetilde \psi}_{lp}(\Delta {\bf k})$, leads to azimuthal asymmetries. For example, on the transverse plane
$(x,y)$, normal to the pump laser,
 \begin{eqnarray}
 P_{oe,x,y}&=& \left[ \cos \theta_c\,\left(d_{15}\,\cos \phi_c\,\sin \theta_{cr}^{\prime}\,\sin \phi_{cr} -
     \cos \theta_{cr}^{\prime}\,\left({d_{22}}\,\cos \phi_{cr}\,\sin \phi_c\,\sin \phi_{cr}^{\prime} +
       {d_{11}}\,\cos \phi_c\,\cos \phi_{cr}^{\prime}\,\sin \phi_{cr}  \right)  \right) ,\right.\nonumber\\
 && \left. {d_{22}}\,\cos \theta_{cr}^{\prime}\,\cos \phi_c\,\cos \phi_{cr}\,\sin \phi_{cr}^{\prime} +
   \left( -\left( {d_{11}}\,\cos \theta_{cr}^{\prime}\,\cos \phi_{cr}^{\prime} \right)  + {d_{15}}\,\sin \theta_{cr}^{\prime} \right)
      \,\sin \phi_c\,\sin \phi_{cr},0   \right] \:\:, \end{eqnarray}
\begin{eqnarray}
\label{asymm} \left(  \int_0^{2 \pi}P_{oe,x,y} d\phi_{cr}^{\prime} \right)^2&=&4 d_{15}^2 \pi^2 \sin^2 \theta_{cr}^{\prime}
\sin^2 \phi_{cr} \left(\cos^2 \theta_c \cos^2 \phi_c+\sin^2 \phi_c \right)\:\:.
\end{eqnarray}
The (simplified) azimuthal dependence indicated by Eq.~(\ref{asymm}) shows a variable intensity for the SPDC rings (See
experimental result in \cite{Otavia}).

\subsection{Modulation by external fields}

Extending our knowledge  about the microscopic behavior of the light-matter interactions in SPDC would help us to examine
other possibilities to use entangled photon pairs. External generalized fields can be added to allow modulations to be
applied to these systems.  Pressure, electric and magnetic fields are natural candidates to exert different modifications on
the optically nonlinear medium. The virtual character of the SPDC process do not allow direct access to energy levels
involved ($\Delta t \rightarrow 0$ leads to $\Delta E\rightarrow \infty$) but, nevertheless, it does not exclude one to
observe important effects related to these virtual processes. For example, a magnetic field may modify electronic levels
with detectable effects on SPDC (Type I or Type II). Induced defects can also be used to probe for local symmetry variations
in crystals \cite{Hong}.
 Several tools can be used to explore this fundamental problem of OAM transfer by a non-linear medium and
may lead to a better understanding of the underlying microscopic physics. Future quantum applications of OAM entanglement,
such as quantum computation or teleportation, may depend on a deep understanding of these OAM transfer process to achieve a
very efficient use and control of quantum entanglements.

\section{Conclusions}

Explicit calculation of the equations determining SPDC processes when OAM is involved were provided for crystals of uniaxial
symmetry. The light-matter non-linear polarizability components $ A_{i,({\bf k},s;{\bf k}^{\prime},s^{\prime})}$ were
calculated giving the complete angular dependence in the crystal reference system as well as in the laboratory system. The
Fourier transform ${\widetilde \psi}_{lp}(\Delta {\bf k})$ was analytically calculated and approximations used were
discussed. Phase matching conditions were obtained and it was shown that spatial transverse coincidence structures can be
calculated. In Ref.~\cite{hugo-barbosa} it was shown that whenever $|F({\bf k}_s,{\bf k}_i)|^2$ (see
Eq.~(\ref{state_vector_final_form})) lacks azimuthal symmetry, an expansion of  $F({\bf k}_s,{\bf k}_i)$ in terms of the
azimuthal angles reveals that --despite the energy conservation condition-- the initial orbital angular momentum $l$
connected with a photon in the incoming mode may not be completely transmitted to the SPDC pair. The transfer may be
partial, with the signal and idler carrying the OAM value  $l^{\prime} \neq l$. The obtained equations for $ A_{i,({\bf
k},s;{\bf k}^{\prime},s^{\prime})}$ and ${\widetilde \psi}_{lp}(\Delta {\bf k})$ allows one to make these expansions to
study specific cases. It is expected that the explicit treatment of the probability amplitude given in this work may allow
further developments in the study of quantum images in SPDC process involving OAM. Analytical tools are then provided to
indicate whether a specific SPDC process may or may not transfer OAM to the conjugate photons. Straightforward extensions of
this work can also be done such as to obtain output profiles in second harmonic generation where one or both of the input
beams are in OAM states. The symmetry properties of the light-matter interaction will be revealed by the up-converted beam.

\section{Acknowledgments}

This work was supported by the U. S. Army Research Office Multidisciplinary University Research Initiative Grant No
W911NF-05-1-0197 on Quantum Imaging.


\section{Appendix}
Expansion of the vector products give the cases $(o,o)$, $(e,e)$ , $(o,e)$ and $(o,e)$:
\begin{eqnarray}{\small
                   \left(  \begin{array}{c}
                         e_{1}e_{1}^{\prime} \\ e_{2} e_{2}^{\prime}\\e_{3}e_{3}^{\prime}\\
       e_{2}e_{3}^{\prime}+e_{3}e_{2}^{\prime}\\  e_{1}e_{3}^{\prime}+e_{3}e_{1}^{\prime}\\  e_{1}e_{2}^{\prime}+e_{2}e_{1}^{\prime}
\end{array}  \right)_{o,o}=
\left(  \begin{array}{c}
                         e_{o,1}e_{o,1}^{\prime} \\ e_{o,2} e_{o,2}^{\prime}\\e_{o,3}e_{o,3}^{\prime}\\
       e_{o,2}e_{o,3}^{\prime}+e_{o,3}e_{o,2}^{\prime}\\  e_{o,1}e_{o,3}^{\prime}+e_{o,3}e_{o,1}^{\prime}\\  e_{o,1}e_{o,2}^{\prime}+
       e_{o,2}e_{o,1}^{\prime}
\end{array}  \right)=
\left(  \begin{array}{c}
                         \sin \phi_{cr} \sin \phi_{cr}^{\prime}\\  \cos \phi_{cr} \cos\phi_{cr}^{\prime}\\0\\
      0\\ 0\\ -\cos \phi_{cr} \sin \phi_{cr}^{\prime}-\sin \phi_{cr} \cos \phi_{cr}^{\prime}
\end{array}  \right)\:,}
\end{eqnarray}
\begin{eqnarray}{\small
                   \left(  \begin{array}{c}
                         e_{1}e_{1}^{\prime} \\ e_{2} e_{2}^{\prime}\\e_{3}e_{3}^{\prime}\\
       e_{2}e_{3}^{\prime}+e_{3}e_{2}^{\prime}\\  e_{1}e_{3}^{\prime}+e_{3}e_{1}^{\prime}\\  e_{1}e_{2}^{\prime}+e_{2}e_{1}^{\prime}
\end{array}  \right)_{e,e}=
\left(  \begin{array}{c}
                         e_{e,1}e_{e,1}^{\prime} \\ e_{e,2} e_{e,2}^{\prime}\\e_{e,3}e_{e,3}^{\prime}\\
       e_{e,2}e_{e,3}^{\prime}+e_{e,3}e_{e,2}^{\prime}\\  e_{e,1}e_{e,3}^{\prime}+e_{e,3}e_{e,1}^{\prime}\\  e_{e,1}e_{e,2}^{\prime}+
       e_{e,2}e_{e,1}^{\prime}
\end{array}  \right)=
\left(  \begin{array}{c}
                         \cos \theta_{cr}\cos \theta_{cr}^{\prime}  \cos \phi_{cr}\cos \phi_{cr}^{\prime}\\
                           \cos \theta_{cr}\cos \theta_{cr}^{\prime}  \sin \phi_{cr}\sin \phi_{cr}^{\prime}\\
                          \sin \theta_{cr} \sin\theta_{cr}^{\prime}\\
      -\sin \theta_{cr} \cos \theta_{cr}^{\prime}\sin \phi_{cr}^{\prime}   -\cos\theta_{cr} \sin \phi_{cr} \sin \phi_{cr}^{\prime}\\
      -\cos \theta_{cr} \cos \phi_{cr}  \sin \theta_{cr}^{\prime}   -\sin\theta_{cr} \cos \theta_{cr}^{\prime} \cos \phi_{cr}^{\prime}\\
      \cos\theta_{cr} \cos\theta_{cr}^{\prime}  (  \cos\phi_{cr} \sin\phi_{cr}^{\prime}+ \sin\phi_{cr}  \cos\phi_{cr}^{\prime})
\end{array}  \right)\:,}
\end{eqnarray}
\begin{eqnarray}{\small
                   \left(  \begin{array}{c}
                         e_{1}e_{1}^{\prime} \\ e_{2} e_{2}^{\prime}\\e_{3}e_{3}^{\prime}\\
       e_{2}e_{3}^{\prime}+e_{3}e_{2}^{\prime}\\  e_{1}e_{3}^{\prime}+e_{3}e_{1}^{\prime}\\  e_{1}e_{2}^{\prime}+e_{2}e_{1}^{\prime}
\end{array}  \right)_{e,o}=
\left(  \begin{array}{c}
                         e_{e,1}e_{o,1}^{\prime} \\ e_{e,2} e_{o,2}^{\prime}\\e_{e,3}e_{o,3}^{\prime}\\
       e_{e,2}e_{o,3}^{\prime}+e_{e,3}e_{o,2}^{\prime}\\  e_{e,1}e_{o,3}^{\prime}+e_{e,3}e_{o,1}^{\prime}\\  e_{e,1}e_{o,2}^{\prime}+
       e_{e,2}e_{o,1}^{\prime}
\end{array}  \right)=
\left(  \begin{array}{c}
                         -\cos \theta_{cr}  \cos \phi_{cr}    \sin \phi_{cr}^{\prime}  \\
                           \cos \theta_{cr}\sin \phi_{cr} \cos \phi_{cr}^{\prime}\\
                          0\\
      -\sin\theta_{cr} \cos\phi_{cr}^{\prime}\\    \sin\theta_{cr}   \sin\phi_{cr}^{\prime} \\
      \cos \theta_{cr} \cos \phi_{cr}  \cos \phi_{cr}^{\prime}   -\cos\theta_{cr} \sin\phi_{cr} \sin \phi_{cr}^{\prime}

\end{array}  \right)\:,}
\end{eqnarray}
\begin{eqnarray}{\small
                   \left(  \begin{array}{c}
                         e_{1}e_{1}^{\prime} \\ e_{2} e_{2}^{\prime}\\e_{3}e_{3}^{\prime}\\
       e_{2}e_{3}^{\prime}+e_{3}e_{2}^{\prime}\\  e_{1}e_{3}^{\prime}+e_{3}e_{1}^{\prime}\\  e_{1}e_{2}^{\prime}+e_{2}e_{1}^{\prime}
\end{array}  \right)_{o,e}=
\left(  \begin{array}{c}
                         e_{o,1}e_{e,1}^{\prime} \\ e_{o,2} e_{e,2}^{\prime}\\e_{o,3}e_{e,3}^{\prime}\\
       e_{o,2}e_{e,3}^{\prime}+e_{o,3}e_{e,2}^{\prime}\\  e_{o,1}e_{e,3}^{\prime}+e_{o,3}e_{e,1}^{\prime}\\  e_{o,1}e_{e,2}^{\prime}+
       e_{o,2}e_{e,1}^{\prime}
\end{array}  \right)=
\left(  \begin{array}{c}
                         -\sin \phi_{cr}  \cos\theta_{cr}^{\prime} \cos \phi_{cr}^{\prime}      \\
                         \cos \phi_{cr}  \cos \theta_{cr}^{\prime}\sin \phi_{cr}^{\prime}\\
                          0\\
      -\cos\phi_{cr}   \sin\theta_{cr}^{\prime}\\    \sin\phi_{cr}   \sin\theta_{cr}^{\prime} \\
 \cos \phi_{cr}  \cos \theta_{cr}^{\prime} \cos \phi_{cr}^{\prime}     -  \sin \phi_{cr} \cos \theta_{cr}^{\prime}\sin \phi_{cr}^{\prime}
\end{array}  \right)\:.}
\end{eqnarray}
For the collinear and degenerate propagation, these vectors give the particular cases described in
Ref.~\cite{MidwinterWarner}.

The nonlinear polarizabilities are obtained in a straightforward way, giving
\begin{eqnarray}
\label{Poo} {\bf P}_{oo}&=&\left[  \sin \phi_{cr}^{\prime}\,\left( {d_{11}}\,\sin \phi_{cr} -{d_{16}}\,\cos \phi_{cr}
\right) +
     \cos \phi_{cr}^{\prime}\,\left( {d_{12}}\,\cos \phi_{cr} - {d_{16}}\,\sin \phi_{cr} \right)  ,\right. \nonumber\\
 &&\left.  \sin \phi_{cr}^{\prime}\,\left( {d_{21}}\,\sin \phi_{cr} -{d_{26}}\,\cos \phi_{cr}   \right)  +
     \cos \phi_{cr}^{\prime}\,\left( {d_{22}}\,\cos \phi_{cr} - {d_{26}}\,\sin \phi_{cr} \right)  ,\right. \nonumber\\
  &&\left.   \sin \phi_{cr}^{\prime}\,\left( {d_{31}}\,\sin \phi_{cr} - {d_{36}}\,\cos \phi_{cr}   \right)  +
     \cos \phi_{cr}^{\prime}\,\left( {d_{32}}\,\cos \phi_{cr} - {d_{36}}\,\sin \phi_{cr} \right)  \right]\:,
\end{eqnarray}
\begin{eqnarray}
\label{Pee} {\bf P}_{ee}=\left[P_{ee,xcr}, P_{ee,ycr},P_{ee,zcr} \right]\:,\:\:\mbox{where}\hspace{5.7cm}
\end{eqnarray}
\begin{eqnarray}
P_{ee,xcr}=\sin \theta_{cr}^{\prime}\,\left( {d_{13}}\,\sin \theta_{cr} -
        \cos \theta_{cr}\,\left( {d_{15}}\,\cos \phi_{cr} + {d_{14}}\,\sin \phi_{cr} \right)  \right)   +
    \cos \theta_{cr}^{\prime}\,\left( - \sin \theta_{cr}\,
           \left( {d_{15}}\,\cos \phi_{cr}^{\prime} + {d_{14}}\,\sin \phi_{cr}^{\prime}   \right) \right.\nonumber \hspace{3mm}\\ \left.
           +
        \cos \theta_{cr}\,\left( {d_{11}}\,\cos \phi_{cr}^{\prime}\,\cos \phi_{cr} +
           {d_{16}}\,\cos \phi_{cr}\,\sin \phi_{cr}^{\prime}             +         {d_{16}}\,\cos \phi_{cr}^{\prime}\,\sin \phi_{cr} +
           {d_{12}}\,\sin \phi_{cr}^{\prime}\,\sin \phi_{cr} \right)  \right)\:,  \hspace{1.5cm}\nonumber \\
  P_{ee,ycr}=  \sin \theta_{cr}^{\prime}\,\left( {d_{23}}\,\sin \theta_{cr} -
        \cos \theta_{cr}\,\left( {d_{25}}\,\cos \phi_{cr} + {d_{24}}\,\sin \phi_{cr} \right)  \right)  +
     \cos \theta_{cr}^{\prime}\,\left( - \sin \theta_{cr}\,
           \left( {d_{25}}\,\cos \phi_{cr}^{\prime} + {d_{24}}\,\sin \phi_{cr}^{\prime}   \right)  +\right.\nonumber\\\left.
        \cos \theta_{cr}\,\left( {d_{21}}\,\cos \phi_{cr}^{\prime}\,\cos \phi_{cr} +
           {d_{26}}\,\cos \phi_{cr}\,\sin \phi_{cr}^{\prime} + {d_{26}}\,\cos \phi_{cr}^{\prime}\,\sin \phi_{cr} +
           {d_{22}}\,\sin \phi_{cr}^{\prime}\,\sin \phi_{cr} \right)  \right)\: ,\hspace{1.5cm}\nonumber \\
 P_{ee,zcr}=\sin \theta_{cr}^{\prime}\,\left( {d_{33}}\,\sin \theta_{cr} -
        \cos \theta_{cr}\,\left( {d_{35}}\,\cos \phi_{cr} + {d_{34}}\,\sin \phi_{cr} \right)  \right)  +
     \cos \theta_{cr}^{\prime}\,\left( - \sin \theta_{cr}\,
           \left( {d_{35}}\,\cos \phi_{cr}^{\prime} + {d_{34}}\,\sin \phi_{cr}^{\prime}  \right)  +\right.\nonumber\\
           \left.
        \cos \theta_{cr}\,\left( {d_{31}}\,\cos \phi_{cr}^{\prime}\,\cos \phi_{cr} +
           {d_{36}}\,\cos \phi_{cr}\,\sin \phi_{cr}^{\prime} + {d_{36}}\,\cos \phi_{cr}^{\prime}\,\sin \phi_{cr} +
           {d_{32}}\,\sin \phi_{cr}^{\prime}\,\sin \phi_{cr} \right)  \right)\:\:.\hspace{1.5cm}
\end{eqnarray}

\begin{eqnarray} \label{Peo}
{\bf P}_{eo}=\left[P_{eo,xcr}, P_{eo,ycr},P_{eo,zcr} \right]\:,\:\:\mbox{where}\hspace{5.7cm}
\end{eqnarray}
\begin{eqnarray}
P_{eo,xcr}&=& \sin \theta_{cr}\,\left( -{d_{14}}\cos \phi_{cr}^{\prime}   + {d_{15}}\sin \phi_{cr}^{\prime} \right)\nonumber\\
  && + \cos \theta_{cr}\left( \cos \phi_{cr}^{\prime}\,
        \left( {d_{16}}\cos \phi_{cr} + {d_{12}}\sin \phi_{cr} \right)  -
       \sin \phi_{cr}^{\prime}\left( {d_{11}}\cos \phi_{cr} + {d_{16}}\sin \phi_{cr} \right)  \right)\:,\nonumber\\
P_{eo,ycr}&=&\sin \theta_{cr}\, \left( -{d_{24}}\,\cos \phi_{cr}^{\prime}  + {d_{25}}\,\sin \phi_{cr}^{\prime}
\right)\nonumber\\
  && + \cos \theta_{cr}\,\left( \cos \phi_{cr}^{\prime}\,
        \left( {d_{26}}\,\cos \phi_{cr} + {d_{22}}\,\sin \phi_{cr} \right)  -
       \sin \phi_{cr}^{\prime}\,\left( {d_{21}}\,\cos \phi_{cr} +
       {d_{26}}\,\sin \phi_{cr} \right)  \right)\:,\nonumber\\
P_{eo,zcr}&=& \sin \theta_{cr}\,\left( - {d_{34}}\,\cos \phi_{cr}^{\prime}   + {d_{35}}\,\sin \phi_{cr}^{\prime}
\right)\nonumber\\
   &&+ \cos \theta_{cr}\,\left( \cos \phi_{cr}^{\prime}\,
        \left( {d_{36}}\,\cos \phi_{cr} + {d_{32}}\,\sin \phi_{cr} \right)  -
       \sin \phi_{cr}^{\prime}\,\left( {d_{31}}\,\cos \phi_{cr} + {d_{36}}\,
       \sin \phi_{cr} \right)  \right)\:\:.
\end{eqnarray}

\begin{eqnarray}
\label{Poe} {\bf P}_{oe}=\left[P_{oe,xcr}, P_{oe,ycr},P_{oe,zcr} \right]\:,\:\:\mbox{where}\hspace{5.7cm}
\end{eqnarray}
\begin{eqnarray}
 P_{oe,xcr}&=&\sin \theta_{cr}^{\prime}\,\left( -{d_{14}}\,\cos \phi_{cr}   + {d_{15}}\,\sin \phi_{cr} \right)
\nonumber\\ &&+
    \cos \theta_{cr}^{\prime}\,\left( \cos \phi_{cr}\,
        \left( {d_{16}}\,\cos \phi_{cr}^{\prime} + {d_{12}}\,\sin \phi_{cr}^{\prime} \right)  -
       \left( {d_{11}}\,\cos \phi_{cr}^{\prime} + {d_{16}}\,\sin \phi_{cr}^{\prime} \right) \,\sin \phi_{cr}
       \right)\:,\nonumber\\
 P_{oe,ycr}&=& \sin \theta_{cr}^{\prime}\,\left( - {d_{24}}\,\cos \phi_{cr}   + {d_{25}}\,\sin \phi_{cr} \right)
 \nonumber \\&& +
    \cos \theta_{cr}^{\prime}\,\left( \cos \phi_{cr}\,
        \left( {d_{26}}\,\cos \phi_{cr}^{\prime} + {d_{22}}\,\sin \phi_{cr}^{\prime} \right)  -
       \left( {d_{21}}\,\cos \phi_{cr}^{\prime} + {d_{26}}\,\sin \phi_{cr}^{\prime} \right) \,\sin \phi_{cr} \right)\:,\nonumber\\
 P_{oe,zcr}&=& \sin \theta_{cr}^{\prime}\,\left( - {d_{34}}\,\cos \phi_{cr}   +
{d_{35}}\,\sin \phi_{cr} \right)  \nonumber\\ &&+
    \cos \theta_{cr}^{\prime}\,\left( \cos \phi_{cr}\,
        \left( {d_{36}}\,\cos \phi_{cr}^{\prime} + {d_{32}}\,\sin \phi_{cr}^{\prime} \right)  -
       \left( {d_{31}}\,\cos \phi_{cr}^{\prime} + {d_{36}}\,\sin \phi_{cr}^{\prime} \right) \,\sin \phi_{cr} \right)\:\:.
\end{eqnarray}


\begin{thebibliography}{}




%

\bibitem{mair}
A. Mair, A. Vaziri, G. Weihs, and A. Zeilinger,  Nature {\bf 412},
313 (2001).

\bibitem{hugo-barbosa}
H. H. Arnaut and G. A. Barbosa, \prl {\bf 85}, 286 (2000).

\bibitem{sheng_quantph}
S. Feng, C-H. Chen, G. A. Barbosa, and P. Kumar, quant-ph/0703187v1, 20Mar 2007.


\bibitem{WallbornMonken}
S. P. Walborn, A. N. de Oliveira, R. S. Thebaldi, and C. H. Monken, Phys. Rev. A {\bf 69}, 023811 (2004).

\bibitem{MandelWolf}
L. Mandel and E. Wolf, {\em Optical Coherence and Quantum Optics} (Cambridge University Press, New York, 1995). Some
historical references are: D. L. Weinberg, Appl. Phys. Letters {\bf 14}, 32 (1969);
 D. C. Burham and D. L.  Weinberg,
Phys. Rev. Lett. 25, 84-87 (1970);
 Y. B.  Zel'dovich, and D. N. Klyshko,
 JETP Lett. {\bf 9}, 40-43 (1969)
   [translated from Pis'ma Zh. Éksp. Teor. Fiz. {\bf 9}, 69 (1969)];
  W. H. Louisell, A. Yariv, and A. E.  Siegman,
  Phys. Rev. {\bf 124}, 1646-1654 (1961).

\bibitem{PeggPadgett}
D. T. Pegg, S. M. Barnett, R. Zambrini, S. Franke-Arnold, and M. Padgett, New J. of Phys. {\bf 7}, 82 (2005). A simple
azimuthal rotation of the Hamiltonian under the constraint $[J_z,H]=0$ shows azimuthal independence: $H(\Delta \phi)=\exp(i
\Delta \phi J_z )H\exp(-i \Delta \phi J_z )=H+i\Delta \phi[J_z,H]+...=H$.


\bibitem{Hong}
W. Hong and L. E. Halliburton,  K. T. Stevens,  D. Perlov, G. C. Catella, R. K. Route, and R. S. Feigelson, J. Appl. Phys.
{\bf 94}, 2510 (2003).

\bibitem{ShengCharlie}
S. Feng, C-H Chen, G. A. Barbosa \& P. Kumar --to be published-- report
 azimuthal dependences on experimental coincidence patterns in Type-II SPDC.

\bibitem{BW} M. Born, E. Wolf, Principles of Optics (Pergamon Press, New York, 1959) pg. 535.

\bibitem{KramersKronig}
J. D. Jackson, {\em Classical Electrodynamics}, 3rd Edition, Sec 7.10 (John Wiley \& Sons, Inc, 1999).

\bibitem{Eimerl_etal}
D. Eierl, L. Davis, S. Welsko, E. K. Graham, and A. Zalkin, J. Appl. Phys. {\bf 62}, 1968 (1987).

\bibitem{MidwinterWarner}
J. E. Midwinter and J. Warner, Brit. J. Appl. Phys. {\bf 16}, 1135 (1965).



\bibitem{Otavia}
O. Jedrkiewicz, E. Bambrilla, M. Bache, A. Gatti, L. A. Lugiato and P. Di Trapani, J. Modern Optics {\bf 53}, 575 (2006).






\end{thebibliography}
\end{document}